\documentclass[modern]{aastex631}
\usepackage{amsmath}
\makeatletter
\long\def\frontmatter@title@above{}
\long\def\ltx@foottext#1#2{%
 \begingroup
 \expandafter\ltx@make@current@footnote\expandafter{\@mpfn}{#1}%
 \@footnotetext{#2}%
 \endgroup
}%
\makeatother
\shorttitle{What would it take for dark matter to be warm?}
\shortauthors{Schutz}
\begin{document}

\title{What would it take for dark matter to be literally warm?}
\author[0000-0003-4812-5358]{Katelin Schutz}
\affiliation{Department of Physics \& Trottier Space Institute, McGill University, Montr\'{e}al, QC, Canada}
\email{katelin.schutz@mcgill.ca}

\begin{abstract}
\noindent
Warm dark matter (WDM) has served as a valuable benchmark for constraining small-scale structure in the past decades. In this note, I examine what it would take for dark matter to be warm in the literal sense assumed by that benchmark, i.e., a thermal relic that decoupled while relativistic. Satisfying current constraints on the WDM mass, which approach the $\sim$10~keV scale, requires one of three possibilities. One possibility is that there were $\sim 10^4$ relativistic degrees of freedom in the thermal bath at the time of decoupling, which is far beyond what is available in the Standard Model or plausible extensions of it. An alternative is that there was a period of early matter domination whose entropy injection diluted the relic density. However, in this scenario, the perturbations would have evolved through an expansion history that was different from the radiation-dominated one assumed in deriving WDM transfer functions. The third possibility is a dark sector that was never in thermal contact with the Standard Model and was simply born colder via asymmetric reheating. All of these possibilities rely on strong coincidences, where physics that has nothing to do with WDM happens to provide the exact initial conditions assumed in a WDM cosmology. Meanwhile, WDM is often used as a proxy for models with self-consistent thermal histories that generically predict the suppression of structure formation on small scales in a way that is both quantitatively and qualitatively different from WDM. I therefore advocate for a transition to more expressive parameterizations and simulation-based methods in order to extract more useful information from the wealth of upcoming data. 
\end{abstract}

\section*{} \label{sec:intro}
The cold dark matter (CDM) paradigm has been enormously successful at predicting observables on large cosmological scales. However, the idea that dark matter (DM) might have a free-streaming length that erases structure below some characteristic length scale has a history nearly as long as CDM itself. The earliest quantitative treatments of this idea focused on massive neutrinos as hot dark matter (HDM) candidates. \citet{Bond:1980ha} and \citet{Bond:1983hb} showed that a neutrino-dominated universe produces a characteristic suppression of the matter power spectrum below the neutrino free-streaming scale. The subsequent realization that HDM erases too much structure on small scales, driven by comparing neutrino-dominated simulations against early galaxy surveys \citep{White:1983fcs}, motivated the search for intermediate scenarios, including early phase-space arguments about whether ``warm'' particles could account for the dynamics of dwarf galaxies \citep{Melott:1985ni}. \citet{Bardeen:1985tr} provided the general statistical framework for relating the matter power spectrum to observable structures, and the stage was set for what would become warm dark matter (WDM): a thermal relic with a mass and temperature intermediate between HDM neutrinos and CDM. To quote \citet{Bode:2000gq}, ``WDM is just HDM cooled down.'' By dialing the particle mass from the neutrino scale up to the keV scale, one could interpolate continuously between HDM and CDM, obtaining a one-parameter family of models that smoothly varied the free-streaming length.\enlargethispage{-\baselineskip}

Over the following decades, WDM became one of the most well-studied alternatives to CDM, propelled by a series of theoretical and observational developments. The massive-neutrino precedent had demonstrated that small-scale suppression of structure was a generic signature of light thermal relics that decouple while relativistic. This idea found a theoretical home in the particle physics landscape of the 1990s, when Grand Unified Theories (GUTs) and supersymmetry (SUSY) were widely expected to be confirmed at upcoming colliders. These frameworks predict $\mathcal{O}(100)$s of new degrees of freedom beyond the Standard Model (SM), and several of the predicted particles are natural keV-scale thermal relics, most prominently the gravitino in gauge-mediated SUSY breaking \citep{Pagels:1981ke,Moroi:1993mb,Borgani:1996ag}. The prospect of discovering SUSY and simultaneously explaining DM with a WDM candidate lent urgency to understanding the observational signatures. On the observational side, the ``small-scale structure crisis'' that emerged in the late 1990s and 2000s provided a phenomenological motivation for WDM that was largely independent of the particle physics arguments. $N$-body simulations of CDM predicted far more satellite galaxies than were observed around the Milky Way (the ``missing-satellites problem'' \citealt{Klypin:1999uc,Moore:1999nt}) and predicted sufficiently dense subhalos that should have been too massive to remain dark (the ``too-big-to-fail problem'' \citealt{Boylan-Kolchin:2011qkt}).\footnote{The status of these issues has evolved significantly. For instance, the missing-satellites problem appears to have been resolved with improved observations and modeling of survey selection effects and the galaxy-halo connection \citep{Kim:2017iwr,DES:2019vzn,DES:2019ltu}.} WDM offered an economical solution to these problems by suppressing the matter power spectrum below the free-streaming scale, reducing the abundance and central densities of small halos: fewer low-mass halos form in a WDM cosmology, and the halos that do form are assembled later than their CDM counterparts and consequently have lower concentrations \citep[see e.g.][]{Bode:2000gq,Lovell:2011rd}.\footnote{Another related issue, the ``cusp-core problem,'' concerned dwarf galaxies whose dynamics indicated flatter central density profiles than the cusps predicted by CDM-only simulations \citep{deBlok:2009sp}. WDM was proposed as a solution to this, however WDM does not outperform CDM \citep{Schneider:2013wwa} as the thermal velocities produce cores that are far too small to be dynamically relevant \citep{KuziodeNaray:2009oon,Villaescusa-Navarro:2010lsj,Maccio:2012qf}.} More recently, the detection of an unidentified emission line at 3.5~keV in stacked X-ray spectra of galaxy clusters and the Andromeda galaxy \citep{Bulbul:2014sua,Boyarsky:2014jta} was interpreted by some as a possible signature of decaying sterile neutrino DM at the $\sim 7\,$keV mass scale. Though subsequent analyses have not confirmed the line \citep{Dessert:2018qih,Dessert:2023fen,XRISM:2025lzv}, it illustrated the continued appeal of connecting a WDM mass to a concrete WDM-like particle physics model.

Constraining small-scale structure remains an important goal for the community, and the WDM mass is still among the most commonly reported parameters in studies that constrain the nature of DM using structure formation. As of this writing, the strongest individual probes each bound $m_{\mathrm{WDM}}$ in the range $5.7$--$7.4$~keV. The 5.7~keV lower bound comes from the high-redshift Lyman-$\alpha$ forest \citep{Irsic:2023equ}, the 5.9--6.2~keV bound comes from Local Group satellite counts \citep{Nadler:2025fcv,Liu:2025vhk}, and the 6.5--7.4~keV bound comes from JWST observations of 28 strong lenses \citep{Gilman:2025fhy,Gilman:2026uvq}. Joint analyses of complementary probes report lower bounds in the $6.0$--$9.7$~keV range \citep{Enzi:2020ieg,Nadler:2021dft}. These probes, their modeling systematics, and the prospects for combining them are comprehensively reviewed in \citet{Nadler:2026waz}.

Clearly WDM has served as a valuable benchmark across decades of work, providing a simple, one-parameter family of models against which to develop and calibrate simulations alongside increasingly precise probes of small-scale structure. In this note, I take the benchmark at face value and ask the simple question: what would it take for dark matter to be \emph{literally} warm? Or put another way, is it possible for dark matter to be a thermal relic that decoupled while relativistic while remaining theoretically and observationally self-consistent? As discussed below, there is very little room for a self-consistent theory of WDM to be viable without strong, unexplained coincidences between physically unrelated model parameters. On the other hand, there are a host of self-consistent theories of DM that deviate from CDM, with small-scale signatures (distinctive cutoffs, oscillatory features, enhancements, etc.) that a single WDM mass cannot represent. The remainder of this note makes the first statement quantitative, surveys the self-consistent model space that WDM constraints implicitly stand in for, and outlines practical steps toward reporting constraints in more expressive terms.

\section{Warm dark matter cosmology} \label{sec:review} 
Suppose we take the quoted constraints at face value, as bounds on the mass of a genuinely warm particle. The relic abundance of such a particle then follows from the same calculation as the cosmic neutrino background. An active neutrino species retains the comoving number density it had while relativistic and in equilibrium, but after $e^\pm$ annihilation transfers entropy to the photons, its temperature is $T_\nu \approx (4/11)^{1/3}\,T_\gamma$, leaving $n_\nu = \frac{3}{4}\left({T_\nu}/{T_\gamma}\right)^3 n_\gamma \approx 112~\mathrm{cm}^{-3}$ today (with an additional factor of $3/4$ from Fermi statistics). A nonzero neutrino mass therefore contributes $\Omega_\nu h^2 = m_\nu/(94~\mathrm{eV})$ to the energy budget of the Universe. A fermionic thermal DM candidate $X$ with two spin degrees of freedom that decouples while relativistic is the same calculation with $T_\nu \to T_X$. Its number density today is rescaled by $(T_X/T_\nu)^3 = (11/4)(T_X/T_\gamma)^3$, giving a relic abundance
\begin{equation}
    \Omega_X h^2 = \frac{m_X}{94 \text{ eV}}\frac{11}{4} \left( \frac{T_X}{T_\gamma}\right)^3,
\end{equation}
where $m_X$ is the mass of the particle and $T_X/T_\gamma$ is the ratio of temperatures between $X$ and the SM. Taking $\Omega_\mathrm{DM} h^2 =0.12$ \citep{Planck:2018vyg}, this implies 
\begin{equation}
T_X = 0.16 \left(\frac{1 \text{ keV}}{m_X}\right)^{1/3}\, T_\gamma. \label{ratio} \end{equation} This temperature ratio, combined with the resulting free-streaming scale, is the input to the WDM transfer function that is most widely adopted in the literature, the fitting formula of \citet{Bode:2000gq} (subsequently updated by \citealt{Viel:2005qj}). This fitting function parameterizes the ratio of the WDM to CDM power spectra as $T(k) = [1 + (\alpha k)^{2\nu}]^{-5/\nu}$ with $\alpha$ depending on $m_X$ and cosmological parameters. The most commonly adopted values are $\nu = 1.12$ and $\alpha = 0.049\,(m_X/1~\mathrm{keV})^{-1.11}\,(\Omega_X/0.25)^{0.11}\,(h/0.7)^{1.22}\,h^{-1}$~Mpc, calibrated against Boltzmann codes at $k < 5\,h~\mathrm{Mpc}^{-1}$ \citep{Viel:2005qj}. As current constraints push toward $\sim 10$~keV, however, the corresponding suppression moves to wavenumbers well beyond the range over which the fit was calibrated. \citet{Vogel:2022odl} revisited the exact Boltzmann calculation and found that a spin-$1/2$ thermal relic produces a \emph{colder} transfer function than the standard fit at fixed mass, with corrections that are significant at the mass scales of current and upcoming constraints (they also provide updated fits, including the first for spin-$3/2$ thermal relics). Even within the literal thermal-relic interpretation, then, the fitting function in widespread use is already a source of systematic error at the current constraint frontier. The COZMIC reanalysis of the Milky Way satellite census, which adopted corrected transfer functions alongside updated subhalo mass function modeling, obtained a WDM mass limit $\sim$10\% weaker than the earlier result (going from $6.5 \to 5.9$~keV; \citealt{DES:2020fxi,Nadler:2025fcv}). 

\section{Entropy transfer from relativistic thermal species} \label{sec:gstar}
Since $X$ decoupled while relativistic, the temperature difference with the SM could be acquired by the entropy transfer of species progressively freezing out of the thermal bath as the universe cools, in complete analogy to the neutrino-photon temperature ratio from the SM. However, the number of degrees of freedom at the decoupling temperature $T_D$ needed to achieve this temperature ratio is
\begin{equation}
     \left( \frac{T_X}{T_\gamma}\right)^3 = \frac{4}{11}\left(\frac{10.75}{g_*(T_D)}\right) = 4 \times 10^{-3} \left(\frac{1 \text{ keV}}{m_X}\right) \Rightarrow g_*(T_D) \approx 953  \left( \frac{m_X}{1 \text{ keV}}\right). 
     \label{eq:gstar}
\end{equation}
Therefore, recent constraints on the WDM mass approaching $\sim10$~keV imply that there would have to have been $\mathcal{O}(10{,}000)$ thermal, light degrees of freedom in the early universe at the time of decoupling, far exceeding the $\sim 100$ of the SM. Even the most conservative analyses, which find $m_X \gtrsim 1.9$--$3.1$~keV once the thermal history of the intergalactic medium is aggressively marginalized over \citep{Garzilli:2019qki,Villasenor:2022aiy}, still require $g_*(T_D) \gtrsim 1800$. This point was already noted in \citet{Bode:2000gq}, and \citet{Viel:2005qj} remark that such particles must ``decouple extremely early.'' At the few-hundred~eV WDM masses under discussion at that time, the required $g_*(T_D) \sim \text{few}\,\times 10^2$ was a bit extreme but not completely beyond credibility. The subsequent order-of-magnitude improvement in the mass constraints has pushed the value of $g_*(T_D)$ well past this point. SUSY roughly doubles the SM particle content, but even in its most extended incarnations, it only introduces $\mathcal{O}(100)$s of new degrees of freedom, which is well short of the $\mathcal{O}(10{,}000)$ required by current WDM mass constraints. Constructions that multiply the SM itself, such as the twin Higgs \citep{Chacko:2005pe} or the $N$-copy sectors of $N$-naturalness \citep{Arkani-Hamed:2016rle}, do not help either. Eq.~\eqref{eq:gstar} counts only species whose entropy is ultimately deposited in the SM bath, whereas mirror sectors by construction retain their own entropy. Indeed, their cosmological viability typically requires the entire mirror sector to be colder than the SM \citep{Chacko:2018vss,Harigaya:2019shz}. 

In summary, one way for DM to be literally warm is for the early universe to have contained an enormous relativistic dark sector, roughly two orders of magnitude beyond the SM or its plausible extensions, that is in thermal equilibrium with the SM bath at the time of decoupling. Via the equality in Eq.~\eqref{eq:gstar}, reproducing the observed relic abundance requires $g_*(T_D)$ to take a precise value at different values of $m_X$. Therefore scanning a family of WDM masses in an analysis implicitly scans a family of dark sectors, the size of whose particle content is imposed by hand to reproduce the DM relic abundance. The particle content and mass spectrum of the theory that determines $g_*(T_D)$ have no intrinsic connection to the physics of WDM, making the tight relationship between $g_*(T_D)$ and $m_X$ a strong, unexplained coincidence. 

\section{Entropy transfer from a period of early matter domination} \label{sec:emd}
To be more economical with the matter content, it is possible to invoke another mechanism that dilutes the relic entropy relative to the SM after it decouples: an early matter-dominated era (EMDE) followed by reheating. This dilution mechanism has a dedicated literature, from gravitino and sterile neutrino realizations to general hidden sectors \citep{Baltz:2001rq,Asaka:2006ek,Bezrukov:2009th,Patwardhan:2015kga,Evans:2019jcs}. If a massive, long-lived particle $\phi$ comes to dominate the energy density of the universe after $X$ decouples and subsequently decays, then the entropy injection into the SM bath dilutes the comoving number density of $X$ relative to the SM without affecting the shape of $X$'s momentum distribution (since $X$ is decoupled). Let $S \equiv s_{\mathrm{after}}/s_{\mathrm{before}}$ denote the entropy dilution factor from the decay of $\phi$. The relic abundance of $X$ is reduced by $1/S$, so the constraint from Eq.~\eqref{eq:gstar} generalizes to
\begin{equation} \label{eq:S_required}
    \left( \frac{T_X}{T_\gamma}\right)^3 = \frac{4}{11}\left(\frac{10.75}{g_*(T_D)\, S}\right) \approx 4 \times 10^{-3} \left(\frac{1 \text{ keV}}{m_X}\right) \;\Rightarrow\; S \approx 9  \left(\frac{m_X}{1 \text{ keV}}\right)  \left(\frac{106.75}{g_*(T_D)}\right),
\end{equation}
where in the last step we have normalized $g_*(T_D)$ to its maximum SM value, since the appeal of this mechanism is that it avoids introducing large numbers of new thermal degrees of freedom. The SM maximum $g_*(T_D)$ therefore represents the best-case scenario, and any later decoupling only increases the amount of dilution that is required. Within the SM, $g_*(T_D)$ is a known function of the decoupling temperature, bounded between $\approx 10.75$ (decoupling at $T_D \sim$~few~MeV) and $106.75$ (decoupling above the electroweak scale). 

The required entropy dilution therefore lies in a finite range, $S \approx 9\text{--}90 \times (m_X/1~\mathrm{keV})$, with larger dilution required the later $X$ decouples. This can be achieved if $\phi$ dominates the energy density at temperature $T_{\mathrm{dom}}$ and decays when the reheat temperature is $T_{\mathrm{RH}} \sim \sqrt{\Gamma_\phi M_{\mathrm{Pl}}}$, yielding an entropy dilution factor of $S \sim {T_{\mathrm{dom}}}/{T_{\mathrm{RH}}}$ in the sudden-decay approximation, up to $\mathcal{O}(1)$ corrections \citep[e.g.,][]{Scherrer:1984fd,Kolb:1990vq}. Long-lived massive relics that come to dominate the early universe arise in theories beyond the SM, including moduli, heavy gravitinos, and their relatives \citep[e.g.,][]{Coughlan:1983ci,Ellis:1984eq,deCarlos:1993wie,Kane:2015jia,Allahverdi:2020bys}. The reheating temperature is separately bounded from below by the requirement that Big Bang nucleosynthesis (BBN) proceed successfully \citep{Kawasaki:2000en,Hannestad:2004px,Hasegawa:2019jsa}. A recent joint analysis of BBN, CMB, and large-scale structure data constrains $T_{\mathrm{RH}} \gtrsim 6$~MeV largely independently of the decay channel \citep{Barbieri:2025moq}. This implies that to obtain the required dilution, we must impose
\begin{equation} \label{eq:Tdom_lower}
    T_{\mathrm{dom}} \gtrsim 54 \text{ MeV} \times \left(\frac{106.75}{g_*(T_D)}\right) \left(\frac{m_X}{1 \text{ keV}}\right) .
\end{equation}
If $X$ decouples before $\phi$ comes to dominate the energy budget, $T_D > T_{\mathrm{dom}}$, then Eq.~\eqref{eq:Tdom_lower} can be recast as an implicit condition on the decoupling temperature alone. For a 10~keV relic to be consistent with the known temperature dependence of $g_*(T)$, we find that $T_D \gtrsim 0.8$--$1$~GeV, well above the QCD phase transition, where $g_*(T_D) \gtrsim 65$ is within a factor of two of its maximal value. The WDM abundance requirement of Eq.~\eqref{eq:S_required} thus pins the duration of the matter-dominated era ($T_{\mathrm{dom}}/T_{\mathrm{RH}} \sim S$) to within a factor of $\lesssim 2$. Specifically, for reheating near the $T_{\mathrm{RH}} \approx 6$~MeV lower bound, the required dilution for a 10~keV relic is $S \approx 90$--$150$ (Eq.~\eqref{eq:S_required} with $g_*(T_D)$ between $106.75$ and $\approx 65$), confining the onset of matter domination to the narrow window $T_{\mathrm{dom}} \approx S\,T_{\mathrm{RH}} \approx 0.5$--$0.9$~GeV. This temperature range becomes correspondingly narrower for larger reheat temperatures, since its fractional width reflects how much $g_*(T_D)$ can vary across the allowed range of decoupling temperatures. For a 10~keV relic with $T_{\mathrm{RH}} \gtrsim 2$~GeV, decoupling is pushed above the electroweak scale where $g_*$ no longer varies, and the window collapses to the single value $T_{\mathrm{dom}} \approx 9\,(m_X/1~\mathrm{keV})\,T_{\mathrm{RH}}$. Alternatively, $X$ may remain coupled at the onset of matter domination and decouple during the EMDE, in which case only the entropy released after decoupling dilutes it. This configuration expands the window for how long matter domination can last, but at the expense of imposing a strong coincidence on the relationship between the decoupling temperature of $X$ and the reheating temperature. The comoving entropy grows steeply while $\phi$ decays, with $S \sim (T_D/T_{\mathrm{RH}})^{5}$ \citep{Giudice:2000ex,Gelmini:2006pw}, so the required $S \approx 900$ for a 10~keV relic that decouples at $g_*(T_D)=10.75$ is achieved only if $X$ decouples at $T_D \approx (900)^{1/5}\,T_{\mathrm{RH}} \approx 4\,T_{\mathrm{RH}}$. In this situation, the SM-$X$ interaction strength and the decay lifetime of $\phi$, two parameters with no common physical origin, must have a fine-tuned relationship in order for decoupling to land just before reheating. 

Beyond the strong coincidences required to match the mass-temperature relation for WDM, the EMDE does not yield precisely the same phenomenology as WDM. Every WDM constraint in the literature is derived from transfer functions computed assuming perturbations evolve through an uninterrupted radiation-dominated era. In the EMDE scenario, the DM instead evolves through a matter-dominated epoch. The cosmology that produces the assumed initial conditions is therefore not the cosmology in which the fitted transfer function was computed. The practical consequences of this are highly scale dependent. Modes near the free-streaming cutoff are far outside the horizon throughout the EMDE, and superhorizon adiabatic perturbations are conserved through arbitrary changes in the equation of state \citep{Wands:2000dp,Weinberg:2003sw}. Additionally, provided that $\phi$ itself carries adiabatic perturbations (i.e., assuming no curvaton-type isocurvature), its decay is a local process that generates no isocurvature between the bath and the decoupled relic on these scales \citep{Weinberg:2004kr,Racco:2022svs}. These modes therefore evolve approximately as they would in the standard calculation from horizon entry onward. In contrast, modes that are inside the horizon during the EMDE ($k \gtrsim k_{\mathrm{RH}}$) experience fundamentally different gravitational environments in the two histories, which has been explored in the contexts of CDM \citep{Erickcek:2011us}, hidden sectors containing relativistic species \citep{Ganjoo:2022rhk}, and cannibal-dominated sectors \citep{Erickcek:2020wzd,Erickcek:2021fsu}. In the EMDE history, such a mode enters the horizon while $\phi$ dominates, and the growing perturbations of the dominating species ($\delta_\phi \propto a$) hold the gravitational potential constant.\footnote{These statements about the subhorizon modes in question are made in the conformal Newtonian gauge. On subhorizon scales, gauge ambiguities in the density contrast are suppressed by $(aH/k)^2$, so these statements carry no practical gauge dependence.} The DM relic is relativistic during this epoch, since $X$ becomes nonrelativistic only once $T_\gamma \approx \text{few}\times \,(m_X/1~\mathrm{keV})^{4/3}$~keV, well after reheating is complete.\enlargethispage{-\baselineskip} The DM therefore cannot fall into these potential wells like pressureless matter would, but rather behaves like neutrinos, so any initial density pattern in the relic is smeared out~\citep{Bond:1983hb}. Full Boltzmann treatments of EMDE cosmologies with relativistic hidden-sector particles find exactly this erasure, with the relativistic species' perturbations suppressed relative to the corresponding cold-limit solution \citep{Ganjoo:2022rhk}. What survives is the response of a relativistic gas to a quasi-static well, where the DM settles toward the equilibrium value $\delta_X \simeq -4\Phi$.\footnote{This is simply hydrostatic equilibrium for a relativistic, collisionless, phase-mixed gas: balancing $\vec{\nabla} P = -(\rho + P)\vec{\nabla}\Phi$ with $P = \rho/3$ gives $\delta\rho/\rho = -4\Phi$, in complete analogy to photon fluid oscillations in the tightly coupled, baryon-free limit of CMB acoustic dynamics \citep{Hu:1994uz}.} In the radiation-dominated history that the typical WDM transfer function assumes, the same mode instead enters the horizon while radiation dominates and the gravitational potential decays. The EMDE history therefore yields a transfer function at $k \gtrsim k_{\mathrm{RH}} \approx 7\times10^{4}\,(T_{\mathrm{RH}}/6~\mathrm{MeV})~\mathrm{Mpc}^{-1}$, the comoving Hubble scale at reheating \citep{Erickcek:2011us}, that differs from the standard WDM transfer function. While these scales are far smaller than the modes currently constrained observationally, there is a growing program of probes targeting DM structure below the threshold of galaxy formation on precisely these scales, through techniques such as pulsar timing, photometric microlensing near lensing caustics, and gravitational-wave diffraction \citep[see][and references therein]{Bechtol:2022koa}. 

In summary, another way for WDM to be literally warm is for an EMDE to dilute its entropy. However, the viability of this scenario depends on strong coincidences relating physical parameters that have no intrinsic relationship. Specifically, the physics of the dilution (set by the mass, lifetime, and initial abundance of the decaying heavy particle $\phi$) has no connection to the WDM mass or the strength of the interactions keeping the DM in thermal equilibrium with the SM. 
Arriving at the right entropy dilution ratio needed to reproduce the observed relic abundance requires $T_{\mathrm{RH}}$, $T_{\mathrm{dom}}$, and $T_D$ to be squeezed between specific values for different WDM masses, with the duration of matter domination confined to a narrow window. Furthermore, this is not internally consistent with the WDM transfer function used in the literature because the same EMDE that provides the required entropy also alters the transfer function relative to the radiation-dominated prediction on scales inside the horizon at reheating. 

\section{Asymmetric reheating of a decoupled dark sector} \label{sec:asym}
A third possibility is that $X$ is part of a dark sector that was never in thermal contact with the SM, but was born colder through asymmetric reheating and thermalized internally \citep{Feng:2008mu,Adshead:2016xxj}. In other words, the dark sector must contain interactions sufficient to thermalize itself while never equilibrating with the SM. A relic that decouples while relativistic within such a sector is a thermal relic in the literal sense, with $T_X/T_\gamma$ set by the initial conditions of the reheating and the matter content of the dark sector rather than by entropy transfer. This path to having literally warm WDM is arguably the most fine-tuned option of the three, because the final temperature of $X$ is now the outcome of a coincidence among several independent variables. Asymmetric reheating only sets the initial temperature ratio of the two sectors. That ratio then drifts whenever either sector loses a species: dark states that annihilate away between reheating and the decoupling of $X$ heat the dark bath exactly as $e^+e^-$ annihilation heats the photons. Applying the bookkeeping of Eq.~\eqref{eq:gstar} to both sectors, the ratio that enters the relic abundance is
\begin{equation} \label{eq:asym}
    \left(\frac{T_X}{T_\gamma}\right)^3 = \left(\frac{T_{X,\mathrm{RH}}}{T_{\gamma,\mathrm{RH}}}\right)^3\left(\frac{g^{\mathrm{dark}}_{*}(T_{X,\mathrm{RH}})}{g^{\mathrm{dark}}_{*}(T_D)}\right) \frac{4}{11}\left(\frac{10.75}{g_*(T_{\gamma,\mathrm{RH}})}\right),
\end{equation}
where $T_D$ is now the dark-sector temperature at which $X$ decouples from its own sector, and $T_{X,\mathrm{RH}}$ and $T_{\gamma,\mathrm{RH}}$ are the different temperatures of the dark and visible sectors at reheating. The reheating temperature asymmetry and the entropy histories of two sectors must conspire to land on the temperature ratio of Eq.~\eqref{ratio} for a given WDM mass, though none of the relevant factors has any intrinsic connection to $m_X$. 

\section{WDM as an approximation for other models} \label{sec:models}
The upshot from the previous sections is that the literal interpretation of WDM has significant structural issues and carries hidden assumptions about tight relationships between intrinsically unrelated parameters. This implies that whatever a quoted WDM mass constraint is constraining, it is probably not a literally warm particle. What kinds of self-consistent scenarios are actually being constrained by measurements of structure on small scales? 

The sterile neutrino is the DM candidate that is most often discussed in the context of WDM due to its similar effect on structure formation, and it is instructive to examine how this model has fared in light of modern constraints on WDM. Sterile neutrinos acquire WDM-like initial conditions in a way that sidesteps the requirements of the previous Sections entirely, because the production mechanism is non-thermal. In the scenario put forth by \citet{Dodelson:1993je}, sterile neutrinos are produced through oscillations with active neutrinos in the early universe, yielding a momentum distribution that, while not precisely thermal, is similar to a Fermi-Dirac distribution. The standard mapping \citep{Viel:2005qj} relates a thermal WDM mass to an equivalent Dodelson-Widrow sterile neutrino mass via $m_s \approx 4.43 \, (m_{\mathrm{WDM}}/\text{keV})^{4/3}$~keV, assuming the sterile neutrino constitutes all of the DM. For decades, this was the default physical model underlying most WDM constraints. However, the same mixing angle that enables production also causes an unavoidable radiative decay $\nu_s \to \nu_a + \gamma$. The combination of increasingly stringent X-ray non-detections of this decay~\citep{Horiuchi:2013noa,Dessert:2018qih,Roach:2022lgo} and structure formation bounds \citep{DES:2020fxi} has now excluded the entire Dodelson-Widrow parameter space for sterile neutrinos constituting all of the DM.\enlargethispage{-\baselineskip} The Shi-Fuller resonant production mechanism \citep{Shi:1998km} remained viable longer. This mechanism involves a primordial lepton asymmetry that resonantly enhances production \citep{Venumadhav:2015pla}, with two consequences. First, the observed abundance is attained at much smaller mixing angles, suppressing the X-ray decay rate ($\Gamma_{\nu_s \to \nu_a \gamma} \propto \sin^2 2\theta \, m_s^5$) and thereby evading the X-ray bounds. Second, the resonance preferentially populates low momenta, yielding a non-thermal spectrum that is ``colder'' than Dodelson-Widrow production (i.e., with a shorter effective free-streaming length). A given Shi-Fuller $m_s$ therefore maps onto a larger equivalent thermal WDM mass than the same $m_s$ produced via the Dodelson-Widrow mechanism, relaxing the structure-formation bounds \citep{Schneider:2016uqi,Baur:2017stq}. But the canonical Shi-Fuller window has now almost closed as well due to combined X-ray and structure-formation constraints, remaining viable only for extremely large lepton asymmetries \citep{Zelko:2022tgf,Vogel:2025aut,XRISM:2025lzv,Yin:2025xad}.

The theoretical landscape has shifted substantially since the WDM benchmark was established. The SUSY scenarios that originally predicted keV-scale thermal relics have not materialized, and sterile neutrino production mechanisms are either excluded or under severe pressure. In the meantime, the lack of experimental evidence for new physics at the electroweak scale has motivated a much broader exploration of dark sector phenomenology. Hidden sectors with internal interactions or that interact with the SM through portal couplings predict a rich variety of possible signatures in the matter power spectrum. Many of these models produce a suppression of small-scale power that is similar to WDM,\footnote{Some dark sector models actually predict an \emph{enhancement} of power on small scales. These include cosmologies with EMDEs \citep{Erickcek:2011us,Erickcek:2020wzd,Ganjoo:2024mie}, post-inflationary axion scenarios whose order-unity field fluctuations collapse into miniclusters \citep{Marsh:2015xka}, and vector DM produced during inflation \citep{Graham:2015rva}. The substructure signatures of such enhancements have now been simulated directly \citep{Nadler:2025crd}. A parameterization like WDM whose single degree of freedom interpolates between ``suppressed'' and ``not suppressed'' cannot express this portion of the model space.} but with qualitatively and quantitatively different transfer function shapes (since the suppression has a different physical origin). It is in this context that the WDM benchmark is best understood: not as a literal hypothesis about the DM, but as a proxy for a broad class of models. 

For example, DM that scatters with baryons suppresses perturbations by collisional damping and diffusion rather than free-streaming; depending on the strength and velocity dependence of the cross section, DM-baryon scattering can even imprint dark acoustic oscillations \citep{Boehm:2000gq,Boehm:2014vja,Dvorkin:2013cea,Gluscevic:2017ywp,Boddy:2018kfv}. Analogous dark acoustic oscillations arise when DM couples to dark radiation, with the precise shape of the power spectrum encoding the dark-sector coupling, temperature, and decoupling epoch \citep{Feng:2009mn,Cyr-Racine:2012tfp,Buckley:2014hja,Buen-Abad:2015ova}. Realizations with a dark version of atomic physics, such as the mirror twin Higgs, add a step-like suppression with oscillations encoding the twin recombination history \citep{Chacko:2018vss,Bansal:2021dfh}, and the nonlinear consequences of atomic DM for Milky Way-analog subhalos have now been simulated directly \citep{Roy:2023zar,Gemmell:2023trd}. Freeze-in production \citep{Hall:2009bx} yields non-thermal momentum distributions whose shape depends on whether decays or scatterings dominate \citep{Bae:2017dpt,Dvorkin:2019zdi,Dvorkin:2020xga,DEramo:2020gpr,Decant:2021mhj,DEramo:2025jsb}. Decays within a dark sector likewise reshape the power spectrum, with daughter particles inheriting a non-thermal velocity kick (and corresponding free-streaming length) from the decay kinematics \citep{Kaplinghat:2005sy,Cembranos:2005us,Huo:2017vef,Fuss:2022zyt,Heeba:2023bik}. Ultralight (fuzzy) DM suppresses collapse through quantum pressure, with a cutoff steeper than any thermal relic followed by oscillations that reflect the wave nature of the field \citep{Hu:2000ke,Marsh:2015xka,Hui:2016ltb}. Cannibal sectors with number-changing $3\to2$ interactions alter the free-streaming history through self-heating \citep{Carlson:1992fn,Hochberg:2014dra}, and even a percent-level cannibal subcomponent alongside CDM leaves a measurable step-like suppression \citep{Buen-Abad:2018mas}. Dynamical dark sectors with multiple unstable states produce multi-modal phase-space distributions whose transfer functions carry several distinct features at once \citep{Dienes:2011ja,Dienes:2020bmn,Dienes:2026prl}.  

None of the examples listed above is literally warm, and in each case the non-WDM shape of the transfer function encodes the unique microphysics of the model. Compressing the information contained in the transfer function to a single WDM mass eliminates the information that distinguishes between different models. Furthermore, recasting a WDM limit for these models requires choosing which feature of the transfer function to match, which can affect the quoted strength of the resulting limit \citep{Konig:2016dzg,Murgia:2018now}. As observational probes reach the precision needed to measure the \emph{profile} of the suppression rather than just its presence, this diversity of shapes is not a nuisance but an opportunity.

\section{Beyond a one-parameter benchmark} \label{sec:beyond}
WDM has clearly been a useful benchmark. However, it is prudent to examine whether this benchmark will continue to serve the field in the coming years. The diversity of well-motivated models described in the previous Section means that a single transfer function shape cannot faithfully represent the space of possibilities. Additionally, the quality and quantity of observational data are reaching the point where the shape of the suppression carries useful information beyond just a suppression scale. Replacing the WDM mass with a different single parameter, like the half-mode wavenumber, would not address the fundamental limitation: a one-parameter family of curves cannot capture a diversity of transfer function shapes.
I therefore advocate for a more nuanced approach depending on the complexity that a given analysis can support.

Ideally, constraints should be reported in terms of a parameterization that captures the shape of the suppression, not just its scale. \citet{Murgia:2017lwo} proposed a three-parameter transfer function of the form $T(k) = [1 + (\alpha k)^\beta]^\gamma$, where $\alpha$ sets the suppression scale, $\beta$ the sharpness of the transition to suppression, and the product $\beta\gamma$ the asymptotic slope at high $k$. This is precisely the WDM fitting form discussed above with the fixed exponents promoted to free parameters, and it was introduced on the grounds that large classes of non-cold models are poorly described by the WDM shape. Different DM models correspond to different regions of the $(\alpha, \beta, \gamma)$ parameter space: thermal WDM, Dodelson-Widrow sterile neutrinos, resonantly produced sterile neutrinos, and fuzzy DM all have distinct combinations of $\beta$ and $\gamma$ even when matched at the same half-mode scale. Constraining all three parameters, rather than fixing $\beta$ and $\gamma$ to their WDM values, immediately reveals whether the data prefer a WDM-like cutoff or something qualitatively different. This requires minimal changes to existing analysis pipelines (one simply allows the shape parameters to float rather than fixing them) and makes results more broadly interpretable across the space of DM models, as demonstrated with Lyman-$\alpha$ forest data by \citet{Murgia:2018now}. The limitation of this approach is that it is still purely phenomenological. The parameters $(\alpha, \beta, \gamma)$ do not map straightforwardly onto physical quantities like coupling constants or particle masses, and the functional form may not capture features like oscillations (as in fuzzy DM or dark acoustic oscillation models). \citet{Hooper:2022byl} generalized this parameterization to include a ``step,'' where the suppression saturates at an intermediate plateau rather than continuing to fall, as occurs in scenarios including a mix of CDM and WDM, among others \citep{Boyarsky:2008xj}. However, the same issue remains: these are still phenomenological shape parameters that are not completely general, with no direct map back to the microphysical parameters of any particular model.

Another approach is to constrain physical parameters of the dark sector rather than phenomenological transfer function shapes. For instance, the Effective Theory of Structure Formation (ETHOS) \citep{CyrRacine:2015ihg, Vogelsberger:2015gpr} parameterizes the dark sector by a small set of effective parameters that control the dark acoustic oscillation and collisional damping scales, together with a self-interaction cross section, and that map onto a range of microphysical models. This framework represents significant progress over the WDM benchmark, and its parameters retain a physical interpretation that purely phenomenological shape parameters lack. Its strength is that it provides an effective description of a broad {class} of models, chiefly those in which DM exchanges momentum with a relativistic species, rather than a universal basis for arbitrary transfer functions. On the other hand, several of the models listed in Section~\ref{sec:models} fall outside the class that ETHOS was designed to describe: freeze-in, cannibal/SIMP DM, and fuzzy DM generate their small-scale signatures through mechanisms other than the dark acoustic oscillations and collisional damping encoded in the ETHOS construction, and the ETHOS parameterization assumes a standard radiation-dominated background, placing modified expansion histories (such as the EMDE discussed above) and inelastic processes that change the sector's particle content over time outside its scope as well. 

Looking forward, simulation-based approaches offer the prospect of constraining specific DM models directly from observational data without requiring any explicit benchmark parameterization. On the inference side used to place constraints, simulation-based inference (SBI) learns the likelihood or the posterior directly from simulated data, requiring neither an analytic likelihood function nor a prescribed functional form for the transfer function. SBI can therefore exploit the full shape information present in the data and can potentially differentiate between qualitatively similar DM scenarios. For instance, if two models are degenerate in a one-parameter summary statistic (like the WDM mass) but distinguishable in the full data vector, this framework automatically exploits the distinguishing information. On the modeling side, emulators (Gaussian processes, neural networks, or similar surrogates trained on simulation suites to learn the mapping from physical parameters to observables) make the forward modeling fast and efficient enough for inference to be practical, whether that inference uses an explicit likelihood or SBI. The flexibility of simulation-based approaches is inherited from the possibilities spanned by the simulation suite rather than imposed by a fitting function. If the data prefer a particular transfer function shape, an analysis trained on simulations with similar transfer functions can discover this without having to pre-specify that shape. This flexibility has implications for the odds of a {first} detection of a deviation from CDM, not just on interpreting a detection after the fact: a deviation whose shape lies outside the assumed template family (a step, a plateau, an oscillation) could be absorbed by astrophysical nuisance parameters, so the expressiveness of the analysis affects the sensitivity of the search itself. By the same token, however, any simulation-based analysis is only sensitive to the parameter space spanned by its training suite. The key disadvantage of the approach is the computational cost of building an appropriately expressive training set, with dedicated simulations for different transfer functions. The rate-limiting step is therefore the availability of non-CDM simulation suites spanning a sufficient range of models.

The effort to simulate more non-CDM cosmologies is now well underway. Simulation suites designed from the outset as multi-model training sets are coming online. For instance, COZMIC zoom-ins with warm, fuzzy, and interacting initial conditions already underpin the $m_{\mathrm{WDM}} > 5.9$~keV Milky Way satellite constraint \citep{Nadler:2025fcv}, and DREAMS varies DM and baryonic feedback parameters jointly across thousands of hydrodynamic simulations \citep[like CAMELS before it;][]{Rose:2024xcb,CAMELS:2020cof}. Simulation-calibrated semi-analytic models are keeping pace, with a unified halo mass function now applicable across CDM, WDM, fuzzy DM, and interacting dark sectors \citep{Benson:2026kdp}. The surrounding analysis infrastructure is also becoming more general, and emulator- and SBI-based inference have delivered flagship bounds. The DESI Lyman-$\alpha$ analysis is already emulator-based and blinded \citep{Chaves-Montero:2026hqd}, the strongest Lyman-$\alpha$ forest limit on fuzzy DM came from Gaussian-process emulation of hydrodynamic simulations, within a framework built to generalize across DM models \citep{Rogers:2020ltq,Rogers:2020cup}, and the thermal-history-marginalized Lyman-$\alpha$ forest bound on WDM came from an emulator over $\sim$1000 hydrodynamic simulations \citep{Villasenor:2022aiy}. Current Lyman-$\alpha$ data are even sensitive to oscillatory features beyond a simple cutoff, and reaching that conclusion required running simulations that contain the oscillations \citep{Yuan:2026zlr}. Beyond just the Lyman-$\alpha$ forest, neural SBI techniques have been used to infer the presence of substructure using strong lenses \citep{Wagner-Carena:2022mrn} and stellar streams \citep{Hermans:2020skz}. Indeed, the recent review of \citet{Nadler:2026waz} identifies simulation-based and machine-learning frameworks as the path toward the strongest and most robust constraints from combined probes of structure.

There is a remaining risk that is more subtle than a lack of tools to parameterize non-CDM cosmologies: the list of benchmark models may simply grow from one (WDM) to three or even $N$. Fuzzy and interacting DM are increasingly becoming the default alternatives, while equally self-consistent scenarios (freeze-in and other non-thermal production histories, entropy-diluted relics and modified expansion histories, cannibal and other number-changing sectors, atomic and mirror sectors, and multi-state dark sector ensembles) typically only enter analyses, when they enter at all, through WDM-equivalent recastings (e.g., \citet{DEramo:2025jsb}). A hand-picked list of benchmarks, however modern, reproduces the original problem with WDM but at a larger scale. Emulation and SBI solve the interpolation and inference problems, but not the coverage problem: a simulation-based analysis is only sensitive to the physics that its simulation suites contain. Extensibility should therefore be a priority in designing new analyses and simulation suites. Expressive parameterizations and emulators built to absorb new model classes can prevent the field from analyzing tomorrow's data through the lens of a stagnant model space.

\section{Discussion} \label{sec:discussion}
The WDM benchmark has been genuinely useful for the field, and the substantial body of work built around it should not be understated. That work drove the development of the observational probes that now constitute the state of the art for studying small-scale structure: Lyman-$\alpha$ forest flux power spectrum measurements \citep{Viel:2005qj,Boera:2018vzq,Irsic:2023equ}, strong gravitational lensing anomaly searches \citep{Dalal:2001fq,Gilman:2026uvq}, satellite galaxy analyses \citep{DES:2019vzn,Nadler:2025fcv}, and analyses of perturbations in stellar streams \citep{Banik:2019smi,Nibauer:2025ezn}. That work likewise drove the supporting simulation machinery, from high-resolution hydrodynamic and $N$-body codes \citep[e.g.,][]{Lovell:2011rd,Vogelsberger:2015gpr} to semi-analytic subhalo models \citep[e.g.,][]{Benson:2010kx,Benson:2012su,Nadler:2021dft}. These tools exist in their current, highly refined form in large part because the WDM benchmark provided a common target against which to calibrate and compare. The benchmark served the field well during the era when the primary observational question was binary: is the power spectrum suppressed below some scale, or not?

The field is now entering a different era. As the field continues to pursue a robust first detection of deviations from the CDM power spectrum, the data are becoming capable of answering a richer question: not just \emph{whether} the power spectrum might be suppressed on small scales, but \emph{what a suppression might look like} if one lies within reach, and what its shape would reveal about the physics of the dark sector. Some analyses are already sensitive to the slope of the suppression, not just its scale, and constraints shift quantitatively when the proper transfer functions are used in place of the thermal WDM approximation. More generally, the sensitivity to the shape of the power spectrum is a progression rather than a fixed property of any individual observable: the scales where the Lyman-$\alpha$ forest is now highly precise were only roughly constrained two decades ago, and the smallest scales accessible today, where only large deviations from CDM are currently detectable, are the ones that ongoing observational and analysis programs are working to sharpen. The volume and precision of small-scale structure data are increasing dramatically, with a further leap ahead as the Rubin Observatory expands the census of Milky Way satellites, stellar streams, and strong gravitational lenses. In this context, a one-parameter benchmark that fixes the shape of the transfer function is a bottleneck and \emph{undersells} the value of how probes of small-scale structure can meaningfully constrain an enormous number of models and physical scenarios. The field will need a framework that can absorb and interpret new small-scale structure data in all its richness. WDM, for all its historical utility, is not that framework.
\newpage
\section*{Acknowledgments}
I am grateful to Juna Kollmeier for the discussions that emboldened the writing of this note. I acknowledge useful conversations and correspondence with Adrienne Erickcek, Cara Giovanetti, Manoj Kaplinghat, Mariangela Lisanti, Ethan Nadler, Keir Rogers, Mark Vogelsberger, and Risa Wechsler. I acknowledge support from the Natural Sciences and Engineering Research Council of Canada Subatomic Physics Discovery Grant, from the Canada Research Chairs program, from the CIFAR Global Scholars program, and from the Alfred P. Sloan research fellowships. This work was further supported in part by grant NSF PHY-2309135 awarded to the Kavli Institute for Theoretical Physics (KITP), where this work was initiated.
\bibliography{literally_warm}{}
\bibliographystyle{aasjournal}
\end{document}